# *Fast trigger logic with digitized time information*

E.Imbergamo[a,*], A.Nappi[a,b], A.Papi[b], A.Riccini[a], M.Valdata[a,b]

[a]University of Perugia, Physics Department, I-06100 Perugia, Italy

[b]INFN, Sezione di Perugia, I-06100 Perugia, Italy

**Abstract**

We present a method for the evaluation, at the first level of trigger, of logical conditions with high time resolution, using the digitized times of fast signals in high rate experiments. We describe a dead-time-less implementation of this method on a commercial FPGA. The method offers an excellent solution to the problem of including veto conditions in the first level of trigger for experiments on rare kaon decays.

## 1. Introduction

The work presented here has been prompted by the recent interest in a new generation of experiments searching for rare K decays, such as $K^+ \to \pi^+ \nu \bar{\nu}$ [1] and for $K^0_L \to \pi^0 \nu \bar{\nu}$ [2], for which precise theoretical predictions of the branching ratios exist [3] and even small deviations from the expected values would indicate new physics [4]. The method developed may fit other types of experiments showing similar, or less challenging, features.

On the experimental side, the study of these decays presents obvious difficulties. Their small expected branching ratios (order of $10^{-10}$) requires a high intensity primary beam, with the disadvantage of a correspondingly high rate of particles on detectors. Signal events have no clear positive signature and background rejection must rely heavily on the use of a hermetic system of veto counters, not only at data analysis level but also at the first level of trigger. High rates in the veto counters are an important source of losses, due to the accidental superposition, within the coincidence windows, with signals unrelated to the event of interest.

---

* Corresponding author, e-mail: ermanno.imbergamo@pg.infn.it

Traditionally, veto conditions are applied using coincidence units, where the overlap among logical signals coming from the veto system and those from the other detectors is evaluated and used for triggering. The inputs to the coincidence units are provided by discriminators, generating a logical signal of preset length on the order of nanoseconds or more, when a particle is detected. Such signals are preliminarily aligned in time by delay chains before entering the coincidence units, where overlaps greater than a few nanoseconds are typically recognized. The use of time windows extending over several nanoseconds implies a loss of interesting data due to accidentals in veto counters.

On the other hand, VLSI technologies allow the implementation of single chip devices which are able to produce fast digitized time information with very high time resolution. An example of such technology is the CERN high performance TDC [5], a 32 channels TDC with configurable resolution up to 25 ps per channel, 5 ns dead time, and 40 MHz read-out frequency. Alternative designs, where the time is provided by digital analysis of the output of high frequency (~1GHz) FADC's, have also been proposed [2]. Using digitized hit times, instead of discriminated logical signals, allows the evaluation of veto conditions in sub-nanosecond time windows, hence reducing the loss due to accidentals, without requiring the development of sophisticated analog electronics. Such evaluation can be performed in commercial FPGAs (Field Programmable Gate Arrays), which are fast, integrate a large amount of resources and are re-programmable.

Different trigger and data acquisition architectures have been proposed for rare kaon decay experiments. In the KOPIO proposal [2], which provided the initial motivation for this work, a dead-time-less first level trigger was used to reduce the data acquisition rate from ~20 MHz to less than 1 MHz. Trigger-less data transfer to the online workstations has been considered in other proposals [1,6]. In this second case, in place of a first level trigger, a hardware processing stage, analyzing readout data "on the fly", could provide primitives useful to speed up data selection in the online workstations.

In this paper we describe the development of a design, which can be used as a building block for different applications, capable of time resolutions matching the time resolution of the detectors, and of accepting input data with the full granularity available for the readout. The resolution and granularity are free parameters of the design that can be adapted to the precise requirements of the experiment, thus determining the size of the devices needed for the implementation. For sake of definiteness, our tests have been performed with a choice of the parameters which would match the specifications of the KOPIO experiment.

KOPIO used a kaon beam produced in short bunches, with average momentum of about 800 MeV/c, in which the time distribution of the signals was widened by significant path and momentum fluctuations. Veto windows with ~5 ns full width were needed for sufficient rejection of the background, which, in presence of rates exceeding 70 MHz in the full region covered by veto counters and of residual time structures, yielded veto accidental losses close to 50%. Although coincidence resolutions in the order of 1 ns were sufficient, a requirement of the experiment, motivated by the need to optimize the efficiency to rejection ratio, was the possibility of fine tuning the delays and the coincidence widths, depending on the position of the veto counters and on the average momentum of particles hitting them.

High energy experiments, instead, could benefit of sub-nanosecond resolutions, provided the front end electronics is capable of applying the corrections required to achieve this performance. We have chosen to develop a design that could be extended to this type of applications, although this would require more performing devices than those which were used in our tests.

## 2. System architecture

In this section we present a possible system architecture for the evaluation of fast logical conditions using digitized hit times.

The inputs are digitized hit times, representing the time of particle crossing in a detector channel, provided by the front-end electronics. They are received synchronously at the frequency of the master clock of the experiment and each hit time is coded into an *n_bits* binary word, describing its time within the period of the master clock, called Time Slot (TS) in the following. We allow the possibility of multiple hits in a TS assuming that *m_Hits_TS* fields of *n_bits* are transmitted at each TS, together with information apt to validate each field.

Different choices of parameters are allowed by the architecture. For definiteness, we assume *m_Hits_TS=3* , *n_bits=8* and a 25 ns master clock period, i.e., for each readout channel, a maximum multiplicity of 3 hits/TS, represented by times ranging from 0 to 249 in units of 100 ps. The configuration with all the *n_bits* at '1', representing an out of range hit time, is used to identify the *no hit* condition.

Figure 1 shows a top view of the system architecture for the trigger module. Data received from different input channels run through four stages of preprocessing (*coarse* *correction*, *fine correction*, *resolution degrading*, *time window generation*) in parallel and then are combined to evaluate a preset logical condition (*logic evaluation*).

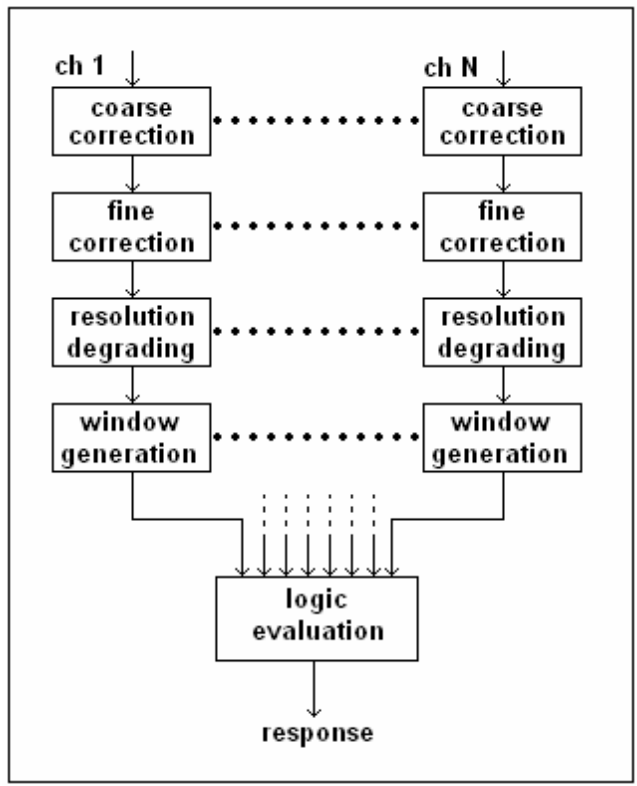

**Figure 1**
**Block diagram of the trigger module.**

The purpose of the first two stages is to perform time alignment, using parameters that compensate different cable and detector delays and optimize the trigger performance. In the first stage a coarse correction, multiple of the TS, is applied by delaying the elaboration of the data channel by a number of clock cycles. A finer alignment, with resolution matching the granularity of input data, is applied in the second stage by adding time offset to each channel data. Both corrections are determined by constant programmable values, different for each channel.

A stage of *resolution* degrading, consisting in the truncation of the least significant $p$ bits, serves to reduce the usage of FPGA internal resources. Including this stage in the trigger allows it to use, if so wanted, the same data sent to the readout, where the number of bits is dictated by the resolution achievable after corrections that can be applied only offline. For example, for the choice of parameters mentioned earlier, with $p=3$, the times processed by the trigger are represented by 5 bits, with a degradation of the resolution from 100 ps to 800 ps. In the figures that illustrate the principles of the algorithms, for clarity sake, we describe the *p*=4 case, in which the TS is subdivided in 16 slices of 1.6ns.

Window *generation* is the key point of the algorithm and consists in changing the representation of the data from a hit time to an array of bits, each representing the state of a logical signal within a time slice submultiple of the TS period. A stream of bits set at '1' starts at the array location corresponding to the hit time and its length is a programmable parameter of the logic which determines the resolution of the coincidences.

The generation of the bit stream is complicated by the fact that, as a consequence of the time alignment and of the extension of the time window, each TS can also receive contributions from the adjacent ones. If the time alignment constants are chosen in such a way that only positive corrections are applied, it is sufficient to keep track of two TS, as illustrated in Fig. 2.

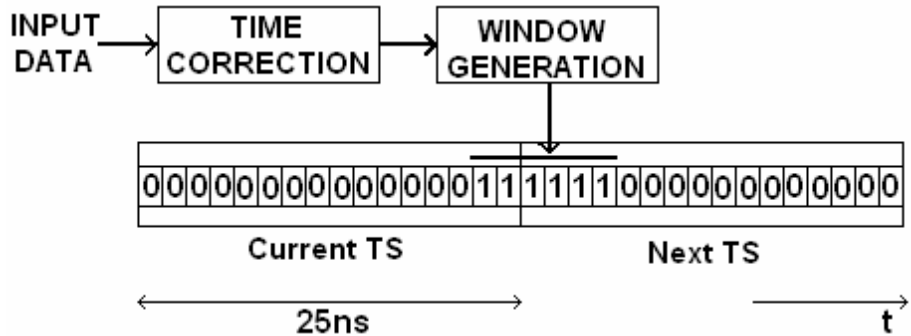

**Figure 2**
**Scheme for the generation of the bit stream. The example shows a hit time occurring at the end of the current TS and extending into the adiacent TS after application of the time width of the signal.**

In order to account for the full activity in one time slot, the array of bits labelled *current time slot* in Fig. 2, must be OR-ed with the array labelled *next time slot* produced at the previous clock cycle. This is achieved by the mechanism illustrated in Fig. 3.

The bit streams of two consecutive time slots are kept in a two word buffer (top line in the figure). At each clock cycle, the fully elaborated current TS is sent to the logic; the next TS becomes the current one, padded with 0's (line 2 in figure) and is ORed with the output of the window generation (line 3), contributing to the current and to the next time slot. The result of the OR (line 4) replaces the two word buffer.

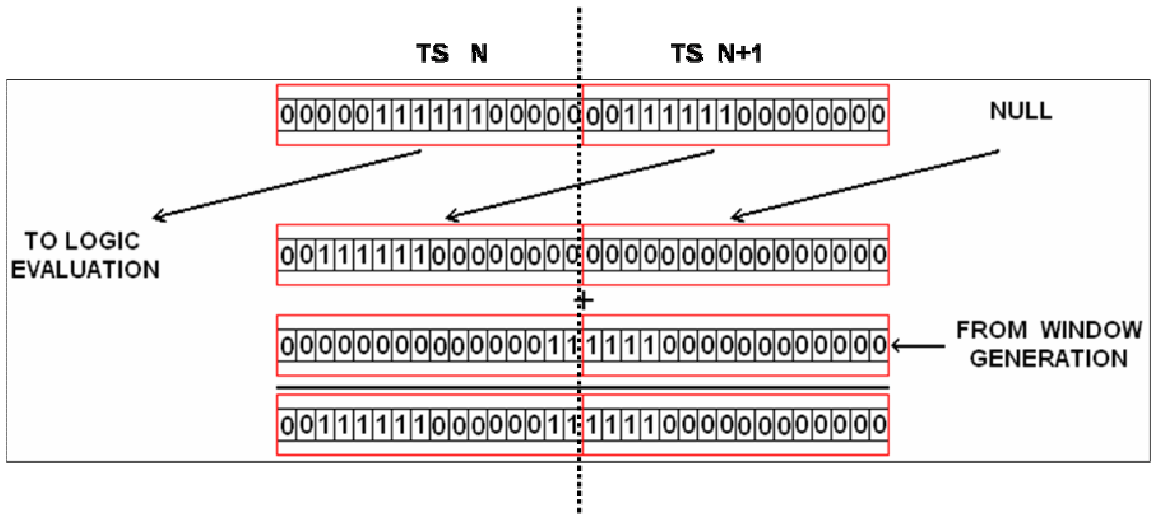

**Figure 3**
**Mechanism for OR-ing consecutive time slots. Bits corresponding to two consecutive time slots, separated by a vertical line, are shown, with the earlier one on the left.**

Finally, the status of signals in all detector channels considered is compared to the preset conditions for all time slices, in the *logic evaluation* block. As illustrated in Fig. 4, the conditions are evaluated in parallel for all time slices and the final trigger response is produced by OR-ing of the results of all time slices.

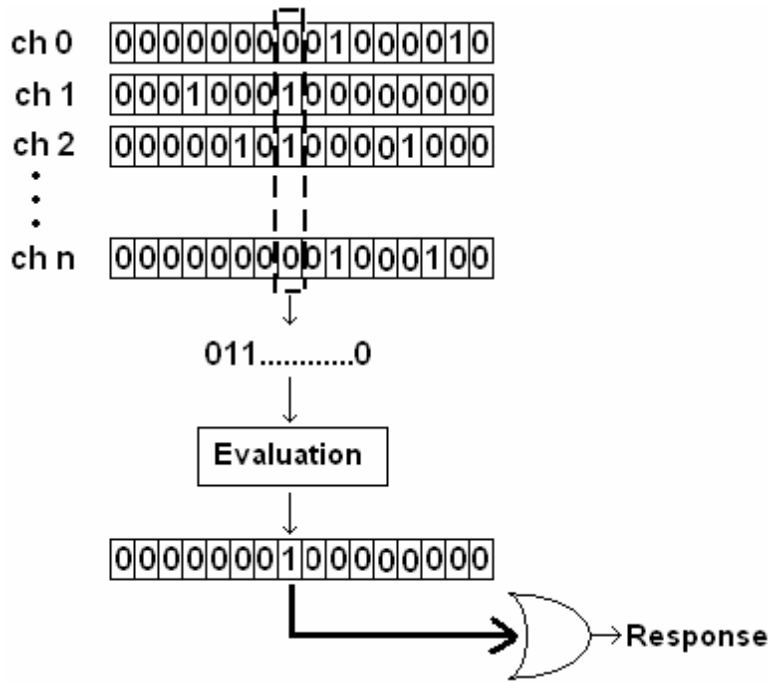

**Figure 4**
**Generation of the trigger response.**

Several possibilities exist for the logic *evaluation* block. In the next section we describe the implementation of a *trigger module*, designed to allow full generality of the logical conditions at fixed latency, using memory look up tables embedded in the FPGA. Such an implementation is not suitable for large number of inputs, since the use of logical resources in the FPGA grows exponentially with the number of inputs.

This would introduce a limitation for applications in rare decay experiments, where veto detectors are usually segmented into a large number of channels, in order to improve double pulse resolution. In this case higher logic densities can be achieved if the requirement of full generality of the logic is dropped, and only 'OR' or 'MAJORITY OR' conditions are used to combine signals from homogeneous sections of the detector. To this end we consider a different type of elaboration which we call "channel reduction" module. It receives hit times from a group of individual channels and returns a "cluster time" for each set of input signals occurring within a preselected window, determined by the time resolution, from each other. A realistic system handling large numbers of inputs can thus be built feeding the "trigger module" described earlier with the outputs of a number of "channel reduction" modules.

A possible scheme for the "channel reduction" module is illustrated in Fig. 5.

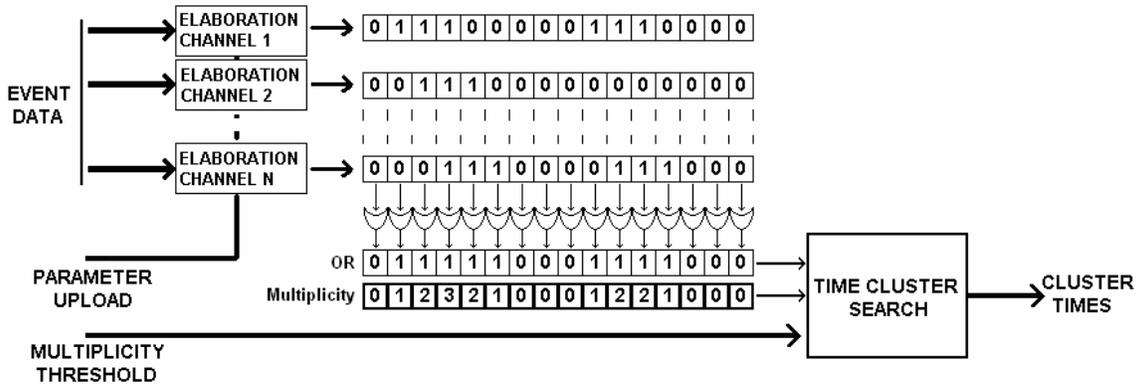

**Figure 5**
**Scheme of the "channel reduction" module.**

Each individual channel is processed as described above to produce an array of bits for each channel. For each time slice, all channels are ORed and the number of channels at '1' is counted. The resulting arrays enter a time cluster search block, which detects streams of consecutive '1' bits and for each stream maps the position of the first '1' bit into the format of a hit time, which will be placed in output depending on the 'multiplicity threshold' requirement.

In the example of Fig. 5, the logical 'OR' among input channels produces two streams of consecutive '1' bits, four and five time slices long respectively. The contents of the *multiplicity* array gives the number of channels hit in each time slice. With a *multiplicity threshold* set at 3, only one bit stream will produce an output cluster time and the *time cluster search* behaves as a 'MAJORITY OR'. With a *multiplicity threshold* set at 1, both the bit streams will produce an output cluster time and the *time cluster search* module behaves as an 'OR'.

As in the example, it may happen that the result of a logical 'OR' among a large number of channels produces more than one "cluster time". For this reason the format of output data has been redefined to accommodate $m\_Hits\_TS \times n\_bits$ per TS. The format of the inputs to the *trigger module* previously described is correspondingly redefined. This choice allows to take advantage of the detector segmentation to achieve a double event resolution better than the double pulse resolution of the individual detector channels, provided the rate within the group of channels is not too high. Posssible modifications of the "time cluster search" stage, allowing to keep double event resolutions even in more extreme rate conditions[1], must be weighted against a possible increase in bandwidth requirements for the output of the "channel reduction" module.

**3. Implementation**

In order to prove the feasibility of the design described in the previous section and to quantify the electronic resources required, we have implemented and tested the "trigger module", using a Microtronix development board equipped with an Altera Stratix® FPGA 1S25F780C6. Although the design is rather general, some of the implementation choices have been tailored to the FPGA used.

Our implementation is fully pipelined. Synchronous inputs from the read-out electronics are processed in subsequent steps and the trigger response is asserted after a predictable number of master clock cycles without introducing dead-time. 5 clock cycles are needed to implement our algorithm. This does not include additional cycles needed to implement the coarse correction, described later.

---

[1] The basic implementation of the "channel reduction" module described suffers of inefficiencies when consecutive events occur separated by an amount of time comparable to the preset length of 1-bit stream, associated to the single hit. In such a case they will appear as only one event and the "time cluster search" module will produce only one cluster time. A possibility for a general solution to this problem is to modify the "time cluster search" stage, in such a way to output as many cluster times as necessary in order to match the overall length of the 1-bit streams. In the example of figure 5, this would lead to 4 cluster times, one starting at the second time slice (ts), a *fake* one starting at the third ts, one starting at the tenth ts and a *fake* one staring at eleventh ts.

Trigger conditions and the parameters for hit time alignment are loaded into internal FPGA memories from a remote PC via a USB interface. Although a hard coded definition of trigger conditions and parameters would lead to a simpler implementation, our choice allows to change them without affecting neither the latency of the system nor the amount of FPGA resources needed.

Hit times from read-out electronics are delivered, via fast serial LVDS links, into differential FPGA I/O pins equipped with serializer/deserializer circuitry. The achievable bandwidth for LVDS transmission depends on the specific target device and exceeds the Gbps/channel in last generation FPGAs. Therefore, with our design parameters the system can receive multiple hit times in the same clock cycle from the read-out (3 hits/TS). After deserialization, data enter the *coarse correction* block. The processing follows the flow of Fig. 6.

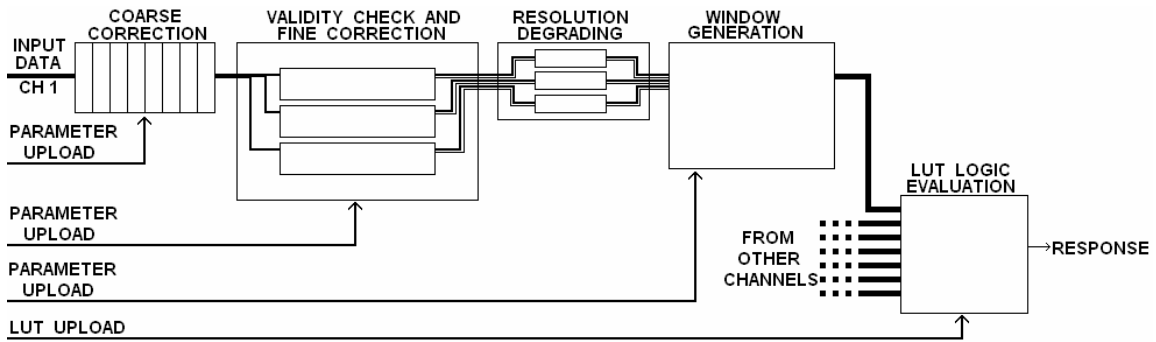

**Figure 6**
**Scheme of the trigger module.**

At each master clock cycle, data from each front-end channel are written in a dual port RAM at a given address and read, after a preset number of cycles. The RAM acts as a delay line with a time granularity of 1 TS. After this *coarse correction*, all hits in the TS are processed in parallel. The time is extracted and a preset positive constant, within the interval 0-249 is added to it (*fine correction*). A comparison with the *no hit* condition is also performed for use in the next stage of logic. The latency introduced by these two blocks is the number of clock cycles that hit times spend in the RAM plus one clock cycle needed for the "fine correction". Finally the $p$ least significant bits are truncated (*resolution degrading*).

After truncation, the data of all hits in the TS enter the *time window generation* block. The mechanism sketched in Fig. 3 is implemented using a $3 \times 2^{n\_bits - p}$ bit long shift register to represent hits and their associated bit streams, with $2^{n\_bits-p}$ bits shifted at any clock cycle. The elder $2^{n\_bits-p}$ bits are passed, after the 'OR' procedure previously described, to the next stage of logic. The intermediate $2^{n\_bits-p}$ bits are updated with the bit streams corresponding to hits occurring in the current TS. The remaining $2^{n\_bits-p}$, although corresponding to future times, are needed to hold bits generated by the possible relocation or extension of hits occurring in the current TS. The logic is organized in three clock cycles.

So far the processing is performed in parallel for all input channels, each one providing a $2^{n\_bits-p}$ bits wide array as input to the "LUT logic evaluation" block. Here the bits are regrouped in words containing one bit per channel, with each word corresponding to a single time slice, as shown in Fig. 4. These words are used to address Look-Up-Tables (LUTs), implemented in embedded RAM memories, preloaded with the trigger responses. In order to avoid dead-time, in one clock cycle the evaluation is performed in parallel for all time slices, so that $2^{n\_bits-p}$ copies of the trigger LUT are instantiated[2].

The last stage of logic performs a logical 'OR' of the responses for all time slices of one TS, which finally marks the TS as accepted[3].

Limits to the size of the design that can be accommodated in commercially available FPGA devices are determined by the resources needed for its implementation. In the device chosen for our tests, an Altera Stratix® FPGA, two kinds of resources have to be considered: the Logic Elements (LE) and the embedded memory. For example, the Stratix® EP1S25 mounted in our development boards has ~25000 LE's and ~2Mb embedded memory. Figure 7 shows the amount of each of these two kinds of resources used in our implementation, as a function of the number of input channels.

---

[2] As we shall see later, RAM usage limits the number of input channels for the system. A solution for reducing the usage of RAM, in favour of an increase in the number of channels, is to read the RAM's several times sequentially in a TS. If *f_TS_master* is the master clock frequency and *f_RAM_max* (within 200-400 MHz for most of the FPGA on the market) is the maximum clock frequency for reading RAMs, the number of LUT copies could be reduced by a factor *f_RAM_max/f_TS_master*.

[3] In this implementation we assume that the trigger only marks the TS for acceptance. In a system where the trigger requires several stages, the output section could be modified to detect clusters of contiguous bits and to provide hit times within the TS, along the lines described for the "channel reduction" module.

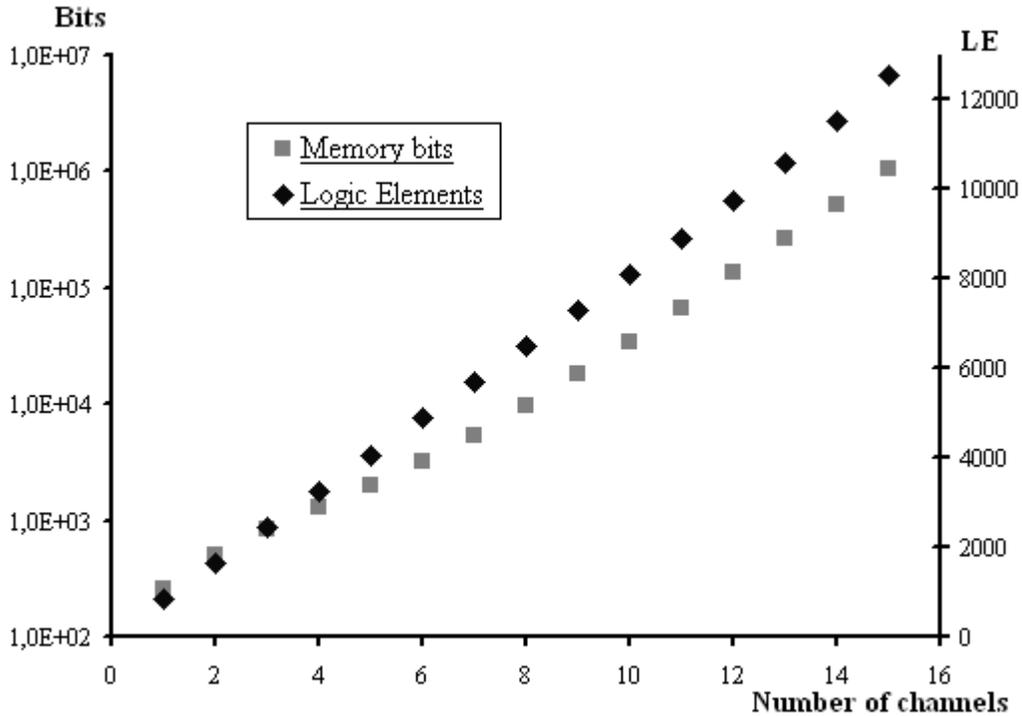

**Figure 7**
**Resources used by the "trigger" module for the configuration with 8 bit input data, 3 bit resolution degrading and maximum multiplicity of hits per channel equal 3. The left scale applies for memory bits, the right one for logic elements.**

The use of embedded memory is mainly determined by the LUT's used in the *logic evaluation* block . This contribution depends exponentially on the number of input channels (determining the size of the LUT for a single time slice) and on the number of bits ( $n\_bits - p$ ) used to represent the time within a time slot (determining the number of time slices, hence of replicas of the LUT). It does not depend on the multiplicity of hits allowed in a time slot.

The use of LE's is mainly determined by the "time window generation" block. It scales exponentially with the number of bits ( $n\_bits - p$ ) used to represent the time within a time slot and linearly with the number of channels and with the multiplicity (*m_Hits_TS*) of hits allowed in a time slot.

For the device used in our tests, the maximum number of channels that we can accommodate is limited to 15 by the embedded memory requirements[4].

---

[4] The maximum number of input channels can be increased to 15 + 8 (where 8 is computed as the ratio of the values of *f_RAM_max* and *f_TS_master* appropriate for our design) if the RAM's are read sequentially in a TS, as explained in footnote 2.

The implementation of LUTs in embedded RAMs looks very limiting, even in higher density devices, due to the exponential growth of the size with the number of channels. We have discarded the alternative choice of implementing the trigger conditions by LEs, that would limit the flexibility in the use of this logic. Changes of the trigger conditions would determine changes in the placement and routing and the functionality of the system might no longer be guaranteed.

For this reason we prefer to address the problem of implementing many read-out channels by introducing a pre-elaboration level of logic which groups hit times from individual read-out channels of a homogeneous detector into a *cluster time*, along the lines presented at the end of the previous section. We have developed a preliminary design of such a *channel reduction* module and simulated its implementation on our device. In this case the limits are determined by the use of LEs. With no particular care in optimization, we have been able to handle 20 channels per module with the parameters of the previous example. We reach 45 when the maximum number of hits per TS in input (*m_Hits_TS*) is limited to one, a more realistic choice for a module receiving its input directly from the front end. These limitations could be enlarged by a more careful optimization.

**4. Design verification**

The design of the "trigger module" and of the "channel reduction" module have been developed and validated by simulations within the Quartus®II development tool. For the "trigger module" a hardware test bench has been set up to test the system at the working frequency.

The test bench makes use of two commercial FPGA development boards, each equipped with a Stratix® FPGA and connected through 8 fast (up to 960Mb/s with short shielded twisted pair cables) LVDS links and a flat cable for exchange of reset/enable controls. One of them is also connected to a PC through a USB interface. The first board simulates an 8 channels detector, while the second one runs the trigger logic and sends the response to the PC.

Offline generated patterns are loaded into a deep FIFO via the USB interface. When the FIFO becomes full, it is emptied at the design frequency of 40 MHz, feeding the trigger board, through the LVDS links, at the nominal frequency. A similar mechanism is used to collect the trigger output onto a FIFO written at 40MHz and emptied slowly through the USB interface. The system has been tested by comparing the response produced by the trigger board with the expected result for a sample of about 20 millions of data patterns generated with a uniform time distribution, without detecting any error.

## Conclusions

We have developed trigger algorithms that evaluate logical conditions with sub-nanosecond time resolution, based on digitized time information from fast detectors. In these algorithms, logic is performed in parallel for thin time slices, using a representation of signals as streams of bits. Arbitrary time resolution, if allowed by the inputs, can be achieved reducing the size of the time slices, with the limit given by the amount of parallelism that can be implemented. Resolutions of fractions of a nanosecond can easily be reached with present day FPGAs.

Realistic systems can be implemented even without resorting to state of the art, expensive devices. A prototype handling 8 input channels at 40MHz with a resolution of 800 ps has been tested on an Altera Stratix® EP1S25. Using the algorithms proposed, a first level trigger for kaon rare decay experiments could be built by a suitable combination of two logic modules, one implementing general logical conditions described in LUTs and another that performs majority logic on signals coming from several homogeneous channels of a large detector.

## Acknowledgements

This work has been performed in the framework of a "PRIN" program financed jointly by Ministero dell'Università e della Ricerca, Italy, by the University of Perugia, Italy and by INFN.

We wish to thank Marco Bizzarri for valuable technical support.

## References

[1] P-326 collaboration, Proposal to Measure the Rare Decay $K^+ \to \pi^+ \nu \bar{\nu}$ at the CERN SPS, CERN-SPSC-2005-013, June 2005.

[2] J.R. Comfort et al., KOPIO Project Conceptual Design Report", http://www.kopio.bnl.gov/cdr2005_nsf.pdf, April 2005

[3] D. Bryman, A. J. Buras, G. Isidori and L. Littenberg, Int. J. Mod. Phys. A21 (2006) 487.

[4] A. J. Buras, F. Schwab and S. Uhlig, Waiting for precise measurements of $K^+ \to \pi^+ \nu \bar{\nu}$ and $K_L^0 \to \pi^0 \nu \bar{\nu}$, hep-ph/0405132, 2004.

[5] J. Christiansen et al., A data driven high performance time to digital converter, in: Proceedings of the 6$^{th}$ Workshop on Electronics for LHC Experiments, Cracow, Poland, 11-15-September 2000, CERN 2000-010, 169-173, 2000; HPTDC web site: http://tdc.web.cern.ch/TDC/hptdc/hptdc.htm

[6] H.Nguyen, The CKM experiment, FERMILAB-CONF-02/254-E, 2002